\documentclass[sigconf,screen,nonacm]{acmart}
\usepackage{mathtools}
\usepackage{graphicx}
\usepackage{dblfloatfix}
\usepackage{multirow}
\usepackage{makecell}
\usepackage{booktabs}
\usepackage[labelfont={bf,it},font+=up,textfont=it,labelsep=period,figurename=Fig.]{caption}
\DeclareCaptionFont{slightlylarger}{\fontsize{9.5}{11.4}\selectfont}
\usepackage{tcolorbox}
\tcbuselibrary{skins,breakable}

\hypersetup{hidelinks}

\newtcolorbox{myregbox}[3][]{
  breakable,
  enhanced,
  colback=#2!5,
  colframe=black!50,
  boxsep=0.5mm,
  #1,
}

\renewcommand{\epsilon}{\ensuremath\varepsilon}
\renewcommand{\phi}{\ensuremath{\varphi}}

\AtBeginDocument{%
 }

\newcommand{\sol}{FissionReady}
\begin{document}

\title{\sol{}: Joint Workload and Power Scheduling for Data Centers Powered by Small Modular Reactors}

\author{Raghavendra Kanakagiri}
\affiliation{
 \institution{IIT Tirupati}
 \city{}
 \country{}
 \texttt{raghavendra@iittp.ac.in}
}

\author{Rohan Basu Roy}
\affiliation{
 \institution{University of Utah}
 \city{}
 \country{}
 \texttt{rohanbasuroy@sci.utah.edu}
}

\author{Yankai Jiang}
\affiliation{
 \institution{Northeastern University}
 \city{}
 \country{}
 \texttt{jiang.yank@northeastern.edu}
}

\author{Pranathi Wuppuluru}
\affiliation{
 \institution{Northeastern University}
 \city{}
 \country{}
 \texttt{wuppuluru.p@northeastern.edu}
}

\author{Devesh Tiwari}
\affiliation{
 \institution{Northeastern University}
 \city{}
 \country{}
 \texttt{d.tiwari@northeastern.edu}
}

\renewcommand{\shortauthors}{Kanakagiri et al.}

\begin{abstract}
Data centers are increasingly exploring small modular nuclear reactors (SMRs) as a carbon-free power source, but variable datacenter demand and negative grid prices require the SMR plant to load-follow rather than run at constant output. Load following is uniquely challenging and complex for SMR plants due to underlying nuclear physics. We propose \sol{}, a datacenter scheduler that tracks each module's fuel age, xenon state, and remaining flexibility to determine per-module operating envelopes, then coordinates load-following, batch deferral, and grid purchasing through a two-timescale optimization. \sol{} achieves zero involuntary shutdowns and zero batch deadline misses while reducing water consumption by roughly one-third and grid cost by roughly half compared to a same-sized base configuration.
\end{abstract}



\maketitle

\section{Introduction}
\label{sec:introduction}

\noindent \textbf{\textit{Background and Motivation.}} Large-scale data centers and HPC sites are increasingly exploring small modular nuclear reactors (SMRs).
In particular, a multi-module SMR plant consisting of several independent reactor modules at a single site can supply hundreds of megawatts of carbon-free electricity consistently, albeit with a significant water footprint, and can continue operating even as individual modules are refueled or taken offline.
But on the demand side, datacenter and HPC power demand (load) is inherently variable.
Batch workloads and online service arrival patterns can be irregular and bursty.
An SMR plant sized for peak demand will produce more power than the datacenter consumes during off-peak hours, leading to electricity wastage, cost inefficiency, and an unnecessarily high water footprint.

The surplus electricity output increases the water footprint, and exporting surplus power to the grid during periods of high renewable generation can actually incur a cost instead of revenue as wholesale prices turn negative.
An SMR plant sized for average demand avoids the surplus problem, but cannot cover the peak load without purchasing grid electricity.
Therefore, \textit{an intuitive solution is for SMR plants to produce electricity in a ``load-following'' manner, adjusting the power output to match the combined pattern of datacenter demand.}

Unfortunately, ``load following'' electricity production is a uniquely challenging problem for SMR plants constrained by nuclear reactor physics .
When a reactor module reduces power production to match reduced load, a fission byproduct (Xenon-135) can build up for several hours and absorb neutrons, making future power maneuvering difficult and potentially unsafe.
A critical insight is that, in a multi-module SMR plant, the physics constraints governing load following differ significantly across modules.
Each module has its own fuel cycle and can be at a different stage at any given time due to staggered refueling, and hence, has different power production maneuvering margin/flexibility due to its xenon buildup history.
Identical current power output does not imply identical future maneuverability.
\textit{Therefore, the focus of this work is identifying a critical and timely gap and proposing a solution:} a job scheduler which is fundamentally aware of reactor physics constraints and can adapt to load changes and coordinate power production and fueling of multi-module SMR plants safely (no shutdown) and efficiently, while being cost-effective and managing water footprint.

\vspace{1.5mm}
\noindent \textbf{\textit{Brief Overview of \sol{}.}} This work proposes \sol{}, a novel job scheduler for datacenters powered by multi-module SMR plants by integrating and systematically exploiting reactor physics knowledge, including each module's fuel age, xenon buildup, recent power production history, and remaining power-production flexibility to respect nonlinear xenon dynamics.

\sol{} designs a novel mechanism to coordinate load-following across SMR modules within a plant, determining when and how much each SMR module should ramp, when to purchase grid electricity, and when to defer batch work -- doing this while safely respecting per-module physics constraints, maintaining workload quality of service, and reducing water consumption.
\sol{} achieves this by introducing \textit{the concept of reactivity headroom}, an abstraction that turns reactor state into a scheduling quantity by comparing each module's compensation ceiling against its predicted future xenon peak, plus a safety margin.
This abstraction is critical since SMR modules at the same power output level can have very different safe futures due to their recent ramp-up and ramp-down history.
Leveraging reactivity headroom estimation, \sol{}'s two-level planner derives a feasible power for each module by solving the nonlinear iodine-xenon ordinary differential equations outside the optimization, then passes linear box constraints to an outer-level planner. At a finer
timescale, an inner scheduler executes the outer planner's decisions, dispatching individual jobs while responding to short-term demand fluctuations (details in Sec.~\ref{sec:scheduler_design}).

\vspace{1.5mm}
\noindent \textbf{\textit{Flagship Results of \sol{}.}} \sol{}'s extensive experimental evaluation, driven by production workload traces and a nuclear physics model, demonstrates that its physics-aware scheduling improves cost, water, and efficiency tradeoffs while preserving service guarantees. Compared with the same-sized baselines and a competitive fixed load-following design, \sol{} safely exploits module-specific flexibility to better match SMR output to datacenter demand, cutting grid electricity cost by roughly half, reducing water consumption by 31\%, and reducing SMR energy waste from 29.8\% to 1.7\%, while preserving low batch wait and zero deadline misses. \textit{Our full framework, including the physics model and scheduling codebase, will be open-sourced to the
community to accelerate future research upon acceptance.}

\section{Background}
\label{sec:background}

The primary metric used to describe reactor behavior (balance between neutron production and loss) is the \emph{effective multiplication factor}, $k_{\text{eff}}$, which measures whether each ``step of the chain reaction'' produces enough new neutrons to keep the chain reaction going. More specifically, it compares the number of neutrons created by fission in one step with the number lost in the previous step through absorption or leakage. If $k_{\text{eff}} = 1$, the reactor is \emph{critical}: the chain reaction ``sustains'' itself, and reactor power generation remains steady. If $k_{\text{eff}}$ is above 1, power generation tends to rise; if it is below 1, power generation tends to fall. In practice, operators usually refer to this in terms of \textit{reactor reactivity}, denoted by $\rho$, which measures how far the reactor is from this critical state~\cite{usdoe_doe_hdbk_1019_v2}. It is defined as
\begin{equation}
\label{eq:reactivity}
\rho = \frac{k_{\text{eff}} - 1}{k_{\text{eff}}}
\end{equation}
Reactivity is often reported in \textit{pcm} (per cent mille), a unit commonly used in reactor engineering, where $1\ \text{pcm} = 10^{-5}$~\cite{nuclear-power_pcm}.

\noindent\textbf{Xenon Poisoning and Excess Reactivity.} Xenon-135 is produced in two ways. Some is created directly by fission, and some comes indirectly from the radioactive decay of Iodine-135 ($^{135}$I). Iodine-135 has a half-life of about seven hours, and Xenon-135 has a half-life of about nine hours. Because these timescales are measured in hours rather than seconds or minutes, xenon does not respond immediately to changes in reactor power. Instead, its concentration evolves gradually.

This delay is what makes xenon especially important during power maneuvers. Iodine-135 acts like a reservoir that builds up while the reactor is operating, roughly in proportion to power.
After power is reduced, the iodine already present in the core continues decaying into Xenon-135, so xenon concentration can keep increasing for several hours even after power has been lowered due to the delayed decay of Iodine-135. The higher the power before a reduction, the larger the iodine inventory, and the more xenon will be generated afterward. When a reactor remains at a steady power level for long enough, the concentrations of $^{135}$I and $^{135}$Xe eventually reach equilibrium where production and loss balance out.

\emph{Load following} is the practice of deliberately raising or lowering reactor power to match a changing electricity demand. It is possible only if operators have enough \textit{controllable reactivity} to maintain reactor criticality throughout the maneuver. \emph{Excess reactivity} is the amount of positive reactivity built into the fuel beyond what is needed to sustain one particular steady operating condition (power generation level). However, this quantity is not fully available for power maneuvering (i.e., changing the power generation level). Much of ``excess reactivity'' is committed to meeting safety and operational requirements that must be met before any power generation level adjustment can be made. Several parasitic demands must be satisfied first.

The \textit{shutdown margin} is the reserve that guarantees the reactor can be made safely subcritical even if the most effective control rod fails to insert. The \textit{temperature defect} is the reactivity consumed in bringing the cold core to hot operating temperature. The \textit{power defect} is the additional change associated with moving from zero to full power.

\vspace{1mm}
As discussed earlier, Xenon-135 is a strong neutron absorber whose concentration depends on reactor power history. When the reactor stays at a fixed power level long enough, iodine and xenon concentrations settle to steady values. The resulting Xenon-135 concentration (\textit{equilibrium xenon}) imposes a continuing negative reactivity penalty that must also be offset to keep the reactor critical. After all of these demands are met, what remains is the available maneuvering margin (excess reactivity), the reactivity reserve that operators can actually use during load following to counter xenon transients while maintaining criticality. A module with more remaining margin has greater flexibility to change power. A module with less margin may still operate safely, but has much less freedom to adjust power without violating reactor-physics limits.

\section{\sol{}: Insights and Design Drivers}
\label{sec:motivation}

\noindent\textbf{Impact of fuel cycle of a module and staggered fueling among different modules.} Recall that Xenon poisoning creates a multi-hour lag between a power change and its full effect on the reactor. Also, recall that each module, in a multi-module plant, has its own fuel cycle and can be at a different stage at a given time from other modules. One may have just been refueled (beginning of cycle, BOC), another may be near spent (end of cycle, EOC), and others may be in between. Different modules may have been assigned different power changes at different times. One module might be at steady-state equilibrium while another is in the middle of a transient from a recent ramp. Even if both modules are currently producing the same power, their xenon concentrations, and therefore their remaining control authority, can differ substantially.

\begin{figure*}[t]
\centering
\includegraphics[scale=0.33]{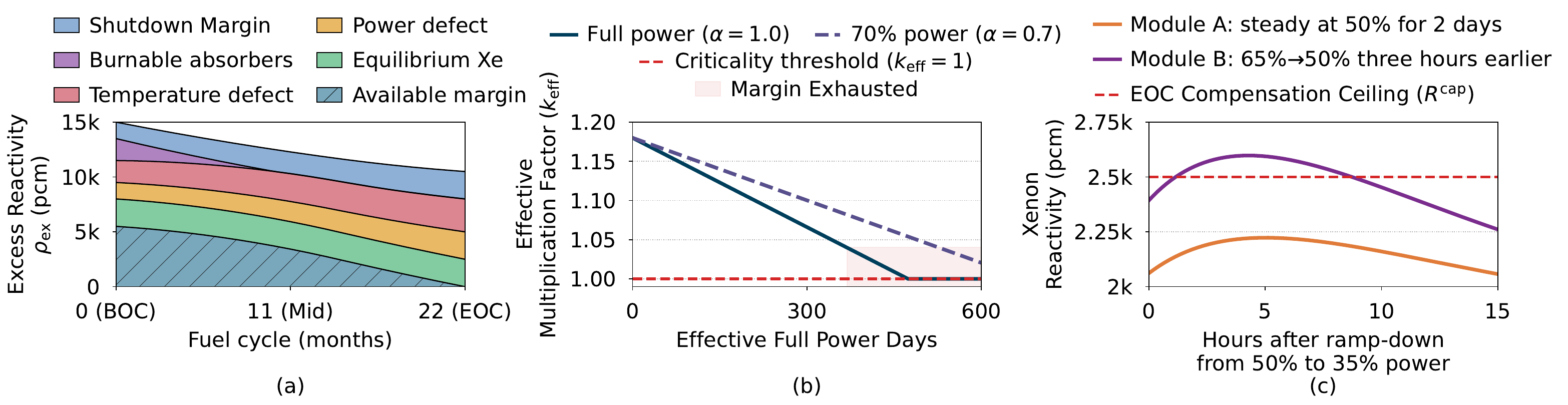}
\caption{Fuel-cycle position and recent power history jointly determine a module’s maneuvering flexibility.
(a) The reactivity budget shrinks across the fuel cycle.
(b) Lower-power operation (100\% capacity vs 70\%) preserves maneuvering margin for longer.
(c) Two end-of-cycle modules at the same current power can still have different feasible futures. The steady module (Module~A, held at 50\% power for two days) remains below the compensation ceiling after a further ramp-down (50\% to 35\%); the recently ramped module B exceeds it because of its elevated iodine-xenon state.
}
\label{fig:reactivity_margin_boc_eoc}
\end{figure*}
\vspace{1.5mm}
Figure~\ref{fig:reactivity_margin_boc_eoc}(a) and (b) illustrate this progression -- how the excess reactivity and available margin for changing the power generation level change over the fuel cycle and capacity factor. Figure~\ref{fig:reactivity_margin_boc_eoc}(a) shows a typical 18 to 24-month fuel cycle with beginning, middle, and end. A reactor's fuel carries more reactivity than is needed for the immediate chain reaction. This excess sustains the reactor over an 18 to 24-month fuel cycle. As fissile atoms are consumed, the excess gradually depletes until the fuel must be replaced.
At the same time, three of the parasitic demands (Section~\ref{sec:background}) on this excess grow. The shutdown margin, temperature defect, and power defect all increase over the cycle.
Burnable absorbers, which suppress excess reactivity at the beginning of the cycle, are consumed by mid-cycle and provide no further benefit.
The result is that the available margin to change power generation level changes over time for any module and depends on position in the fuel cycle.
In a pressurized-water SMR, the available margin for load following shrinks from 5500\,pcm at the beginning to approximately 0\,pcm at the end of the fuel cycle.

Figure~\ref{fig:reactivity_margin_boc_eoc}(b) shows that the capacity factor significantly affects the available margin for changing the power generation level; a lower capacity factor has a much higher available margin for changing the power generation level.

\begin{myregbox}{white}{}
\textbf{Summary.} Staggered refueling places modules at different points in the fuel cycle at any given time. The double squeeze of fuel depletion and growing parasitic demands (parasitic demands are required to operate the reactor safely) means that a module near the beginning of the fuel cycle may have far more maneuvering margin than one near the end of the fuel cycle. Both fuel cycle position and operating capacity factor are critical for a scheduler that attempts to exploit load following.
\end{myregbox}

\vspace{1.5mm}
\noindent\textbf{Impact of fuel cycle history, historic power generation level, and xenon dynamics.} Neither fuel cycle position (fuel age) nor recent power history alone determines whether a module can safely perform a load-following maneuver. These two interact due to xenon dynamics.

To see why, it helps to define a \textit{compensation ceiling} for each module: the maximum total xenon reactivity that the control rods can compensate at the module's current fuel age (position in the fuel cycle).
This quantity can be read from the reactivity budget by subtracting the non-xenon demands (shutdown margin, temperature defect, power defect, burnable absorbers) from the total excess reactivity. It shrinks from roughly 8000\,pcm at BOC to roughly 2500\,pcm at EOC.
Whether a proposed ramp will stay within this ceiling depends on how much xenon accumulates afterward, which is governed by the iodine already present in the core.
The transient is described by the iodine-xenon dynamics~\cite{usdoe_doe_hdbk_1019_v2,choudhury2025physicsinformedunitcommitmentframework}:
\begin{align}
\label{eq:ixe_odes}
\frac{dI}{dt} &= \gamma_I \Sigma_f \phi \;-\; \lambda_I I \notag \\
\frac{dXe}{dt} &= \gamma_{Xe} \Sigma_f \phi \;+\; \lambda_I I \;-\; \lambda_{Xe} Xe \;-\; \sigma_{Xe} \phi \cdot Xe
\end{align}
where $I$ and $Xe$ are the iodine-135 and xenon-135 concentrations, $\gamma_I$ and $\gamma_{Xe}$ their fission yields, $\lambda_I$ and $\lambda_{Xe}$ their decay constants, $\Sigma_f \phi$ the fission-rate density, and $\phi$ the neutron flux (proportional to power). The nonlinear $\sigma_{Xe} \phi \cdot Xe$ term represents xenon destruction by neutron absorption. The iodine and xenon concentrations at any moment determine the future xenon trajectory, so a module's recent ramp history is captured without additional bookkeeping.

Figure~\ref{fig:reactivity_margin_boc_eoc}(c) illustrates this interaction by plotting the xenon trajectories of two end-of-cycle modules after the same 15-point ramp-down from 50\% to 35\% power, but from different recent power histories.
Module~A has been held steadily at 50\% power for two days, so its iodine and xenon inventories are near equilibrium.
Solving Equation~\ref{eq:ixe_odes} forward predicts a peak xenon reactivity of roughly 2220\,pcm, safely below the end-of-cycle compensation ceiling of roughly 2500\,pcm.
Module~B is also currently at 50\% power, but it was at 65\% power three hours earlier. The residual iodine from that higher-power period drives a larger xenon transient after the same ramp-down, causing xenon to peak at roughly 2600\,pcm and exceed the ceiling.
Thus, two modules at the same current power can have different feasible futures because recent ramp history changes the iodine-xenon state.

\begin{myregbox}{white}{}
\textbf{Summary.} Whether a module can safely perform a load-following maneuver depends on both its fuel age (position in the fuel cycle) and its recent power generation level history.
A job scheduler that aims to exploit load-following maneuvers must track both the fuel age and xenon dynamics (iodine and xenon concentrations from recent ramps).
\end{myregbox}

\section{\sol{} Scheduler Design}
\label{sec:scheduler_design}

\sol{} employs an effective design to apply load-following across a multi-module SMR plant that powers a datacenter.
The datacenter runs latency-sensitive online services that must be served immediately and delay-tolerant batch jobs that can be deferred in time to align with favorable grid conditions or available SMR capacity.
\textbf{\sol{} determines when and how much each module should ramp up or down, when to purchase grid electricity, and when to defer batch work.}
\textit{The goal is to do this safely, respecting the per-module physics constraints from the preceding sections, while maintaining workload quality} of service and reducing water consumption.
Batch jobs are delay-tolerant and may run at any point before their deadline with no penalty.

\textit{One of the design challenges is to translate per-module physics constraints (fuel age, xenon dynamics, and the resulting maneuvering limits) into a form that an optimization-based scheduler can reason about and operate on.} We begin by formalizing \sol{}'s key ideas that make
this possible.

\subsection{\sol{}'s Reactivity Headroom Concept}
\label{sec:model}
Section~\ref{sec:motivation} showed that a module's ability to change power depends on its fuel age (position in the fuel cycle), recent power history (recent power generation levels), and the resulting iodine-xenon state. Two modules at the same power and fuel age can have very different feasible futures depending on their current $(I, Xe)$ concentrations (recall Modules~A and~B in Section~\ref{sec:motivation}). To distribute power changes intelligently across a heterogeneous fleet, the scheduler needs to first determine the maneuvering flexibility each module has at any moment.

One of \sol{}'s key ideas is to identify and formally encapsulate this flexibility as \textit{reactivity headroom} of module~$i$, evaluated at a candidate target power~$P$.
\begin{equation}
\label{eq:headroom_def}
{h_{i}(P) = \underbrace{R_i^\text{cap}(b_i)}_{\text{compensation ceiling}} \;-\; \underbrace{\rho_{i}^\text{Xe,peak}(P)}_{\text{predicted peak xenon}} \;-\; \underbrace{\sigma_\text{margin}}_{\text{safety buffer}}}
\end{equation}
$h_{i}(P)$ is the gap between what a module can tolerate and what it will experience. \sol{} quantifies what the module can tolerate as the compensation ceiling $R_i^\text{cap}(b_i)$ and what it will experience as the predicted peak xenon $\rho_{i}^\text{Xe,peak}(P)$. The buffer margin ($\sigma_\text{margin}$) accounts for modeling uncertainty and measurement error.
A module with large headroom is a rich resource. It can absorb deep ramps, take load off more constrained modules, and enable water-saving maneuvers. A module with near-zero headroom is effectively rigid and must hold its current power or risk a shutdown.

\sol{} estimates reactivity headroom using two physics-derived quantities.
First, \sol{} estimates how much xenon the control rods can compensate at the module's current burnup $b_i$. \sol{} obtains this from the reactivity budget by subtracting the non-xenon demands (shutdown margin, temperature defect, power defect, burnable absorbers) from the total excess reactivity.
We denote this quantity as the \textit{reactivity compensation ceiling}, $R_i^\text{cap}(b_i)$, hereafter referred to as the \textit{compensation ceiling} for brevity.

Second, \sol{} needs to predict the worst xenon the module will face if moved to power~$P$ and held there. \sol{} computes this by setting $\phi = \phi_0 \cdot P / P_\text{max}$ in the iodine-xenon ODE (Equation~\ref{eq:ixe_odes}), integrating forward from the module's current concentrations $(I_i(t), Xe_i(t))$ over the entire transient decay window, and recording the peak.
\begin{equation}
\label{eq:xepeak_def}
{\rho_{i}^\text{Xe,peak}(P) \;=\; \max_{\tau \geq t}\; \rho_i^\text{Xe}\!\bigl(\tau;\; \phi = \phi_0 \tfrac{P}{P_\text{max}},\; I_i(t),\; Xe_i(t)\bigr)}
\end{equation}
Because the current concentrations $(I_i, Xe_i)$ already encode the module's full recent power history, no additional bookkeeping is needed. \sol{} uses the worst-case future peak rather than the instantaneous value because a ramp decision made now cannot be retracted. If the committed xenon will exceed the ceiling later, the ramp must be refused now.

\vspace{1.5mm}
\noindent \textit{As flexibility erodes, options shrink.}
\label{sec:consequences}
As $R^\text{cap}$ declines across the fuel cycle, even the shallowest ramp may exhaust a module's headroom.
The minimum feasible power level for module $i$ is the lowest $P$ at which headroom remains non-negative,
\begin{equation}
\label{eq:pmin}
{P^*_{\text{min},i} = \min\;\bigl\{ P \;\big|\; h_i(P) \geq 0 \bigr\}}
\end{equation}
found by binary-searching over the forward-integrated xenon peak (Equation~\ref{eq:xepeak_def}).
Early in the fuel cycle this constraint is slack and the regulatory floor governs.
Near end of the fuel cycle the physics floor may approach 100\%, leaving the module no room to ramp.

\vspace{1.5mm}

\subsection{\sol{}'s Optimization Approach and Architecture}
\label{sec:optimization_approach}
The reactivity constraints formalized above are inherently nonlinear, so the next question is how does \sol{} incorporate them into a tractable optimization while still reasoning about each module individually.

Rule-based heuristics do not naturally capture the multi-hour consequences of xenon transients, aggregate fleet models lose per-module heterogeneity, and reinforcement learning does not by itself provide formal guarantees that every explored action will satisfy headroom constraints. Embedding the iodine-xenon ODE (Eq.~\ref{eq:ixe_odes}) directly into the \sol{}'s planner produces a nonconvex nonlinear program due to the bilinear xenon destruction term.

\vspace{1.5mm}
\noindent \textit{\sol{}'s physics-scheduling decomposition.}
\sol{} decomposes the problem into a \emph{physics subproblem} and a \emph{scheduling master problem}. The physics subproblem solves the nonlinear ODE outside the optimization, computes the current headroom $h_i(p_i)$ per module (Equation~\ref{eq:headroom_def}), and determines the minimum feasible power $P_{\text{min},i}$ by binary-searching for the zero crossing of $h_i(P)$ (Equation~\ref{eq:pmin}). The output is a pair of linear bounds per module, $P_{\text{min},i} \leq p_i \leq P_{\text{max},i}$, encoding the full nonlinear xenon dynamics as simple box constraints.

\sol{}'s scheduling master problem optimizes over these bounds without representing iodine, xenon, or the ODE. \sol{}'s decomposition is tight because the binary search solves the ODE at exact candidate power levels, producing the linear bound consistent with the full nonlinear physics. Because the ODE integration runs independently per module, it is trivially parallelizable, and because the outer planner receives only box constraints, the problem structure remains clean and sparse. At each replanning event, the physics subproblem re-evaluates all module bounds using the latest measured $(I_i, Xe_i)$ states, creating a feedback loop between the physics and the optimization.

\vspace{1.5mm}
\noindent \textit{\sol{}'s Two-level Architecture.}
\sol{}'s scheduler operates at two timescales.
The \emph{inner scheduler} dispatches individual jobs every $\Delta_t$. We refer to each such $\Delta_t$ time slice as an \emph{inner bucket}.
The \emph{outer planner}, implemented as a mixed-integer linear program, re-solves every $N_\text{plan}$ inner buckets.
At each run, \sol{} looks ahead over a rolling horizon of $H$ hours, divided into planning steps of duration $\Delta_T = N_\text{plan} \cdot \Delta_t$.
This lookahead allows the planner to account for xenon transients, batch deadlines, and changing grid conditions beyond the current decision point.

\sol{} decides per-module thermal power setpoints, grid purchases, steam bypass decisions, and batch energy allocation while treating online demand as a committed parameter.
Between replanning events, the inner scheduler and dispatch module operate using the most recent plan.
The outer planner plans the energy budget and the inner scheduler spends it.
\sol{} indexes modules by $i$ and outer planning steps by $k$.

\subsection{\sol{} Scheduling Details}
\label{sec:objective6}

\sol{}'s outer planner minimizes a weighted sum of five terms that capture online-service shortfall, batch deadline misses, SMR-side resource use, grid-side resource use, and wasted energy.
The per-module operating envelopes derived from the headroom formulation (Section~\ref{sec:model}) constrain each module's power, allowing the planner to exploit each module's full available maneuvering flexibility without representing the nonlinear xenon dynamics directly.

{\small
\begin{align}
\label{eq:objective6}
\min \sum_{k} \Big[& \overbrace{(\alpha w_\text{smr} {+} \psi)\! {\textstyle\sum_{i=1}^{N}}\! p_i[k] \Delta_T}^{\text{\footnotesize SMR water+fuel cost}} {+} \overbrace{(\alpha w_\text{grid}[k] {+} \omega\, \pi_\text{floor}[k]) P_\text{grid}[k] \Delta_T}^{\text{\footnotesize grid water+cost}} \nonumber \\[-2pt]
&{+}\; \overbrace{\lambda \, E^{\text{online}}_{\text{unmet}}[k]}^{\text{\footnotesize online SLA}} \;{+}\; \overbrace{\gamma \, E^{\text{batch}}_{\text{deadline\_missed}}[k]}^{\text{\footnotesize batch deadlines}} \;{+}\; \overbrace{\epsilon \, E_\text{waste}[k]}^{\text{\footnotesize energy waste}} \Big]
\end{align}
}

\noindent \textit{SMR-side resource use.}
In Equation~\ref{eq:objective6}, $p_i[k]$ is the thermal power of module~$i$ at planning step~$k$ and $N$ is the number of modules.
The first term penalizes SMR generation through two marginal quantities, water intensity $w_\text{smr}$ and fuel cost $\psi$.
A module in steam bypass still incurs this term, which correctly captures that bypassed output still consumes water and fuel even when it does not serve the datacenter.
We intentionally exclude capital cost and fixed O\&M from the online objective because those costs are sunk over the planner's horizon and do not change the relative ranking of feasible dispatch actions.

\vspace{1.5mm}
\noindent \textit{Grid-side resource use.}
$P_\text{grid}[k]$ is imported electrical power from the grid at planning step~$k$.
Wholesale electricity prices can turn negative during renewable-heavy hours.
If raw negative prices entered the objective directly, a large $\omega$ could create an incentive to buy grid energy purely for arbitrage even when the datacenter has no use for it.
\sol{} therefore clamps price to $\pi_\text{floor}[k] = \max(\pi_\text{min}, \pi[k])$, where $\pi_\text{min}$ represents the minimum transmission and distribution cost that applies regardless of wholesale price.
This preserves a nonnegative monetary signal while still allowing renewable-heavy hours to appear attractive through their lower water intensity $w_\text{grid}[k]$.

\vspace{1.5mm}
\noindent \textit{Service-level and waste terms.}
$E^{\text{online}}_{\text{unmet}}[k]$ is online energy that the fleet and grid together could not deliver and is penalized by $\lambda$, the largest weight in the objective.
$E^{\text{batch}}_{\text{deadline\_missed}}[k]$ is cumulative batch energy whose deadline has passed but that has not been served, penalized by $\gamma$.
$E_\text{waste}[k]$ is surplus energy that the datacenter cannot consume.
Penalizing waste with $\epsilon$ discourages overproduction, makes rod-based reductions preferable to steam bypass, and encourages the planner to absorb unavoidable surplus with batch work.

\vspace{1.5mm}
\noindent \textit{Flexibility induced by the objective.}
The waste penalty $\epsilon \cdot E_\text{waste}[k]$ drives \sol{}'s load-following to match demand by discouraging overproduction.
By itself, however, it does not create an incentive to ramp below demand and substitute low-water grid power.
The water term $\alpha w_\text{smr}$ creates an additional incentive to ramp the SMR below instantaneous demand.
This makes low-water grid substitution attractive during renewable-heavy hours, provided the per-module headroom constraints allow the ramp.
Larger $\omega$ makes grid power more expensive and pushes the planner toward SMR-first operation.
These objective terms define the planner preferences. Next, we define which plans are physically and workload-feasible.

\subsection{Constraints}
\label{sec:constraints6_formulation}
\sol{}'s outer planner does not schedule individual jobs directly.
Instead, it allocates aggregate batch energy across future planning steps while enforcing power balance, deadline feasibility, queue conservation, and per-module operating limits.
The non-deferrable demand in step~$k$ is
\[
E_\text{commit}[k] = (P_\text{idle}[k] + P^{\text{online}}[k] + P^{\text{batch}}_{\text{running}}[k]) \cdot \Delta_T,
\]
where $P_\text{idle}[k]$ is cluster idle power, $P^{\text{online}}[k]$ is online-service demand, and $P^{\text{batch}}_{\text{running}}[k]$ is the energy consumed by batch jobs already dispatched in prior steps. For the current step these quantities are measured and for future steps they are forecast.

\vspace{1mm}
The datacenter does not export surplus, and grid purchases are limited by the datacenter's interconnection, $0 \le P_\text{grid}[k] \le P_\text{grid}^{\max}$, where $P_\text{grid}^{\max}$ is the maximum power that can be imported through the site substation or power-purchase agreement.
The planner enforces aggregate power balance:
\begin{align}
\label{eq:balance6}
{\textstyle\sum_{i=1}^{N}} (p_i[k] {-} q_i[k])\, \Delta_T + P_\text{grid}[k]\, \Delta_T + E^{\text{online}}_{\text{unmet}}[k] \nonumber \\
= E_\text{commit}[k] + E^{\text{batch}}_{\text{budget}}[k] + E_\text{waste}[k].
\end{align}
Here $q_i[k]$ is the portion of module~$i$'s output diverted through steam bypass, so $p_i[k] - q_i[k]$ is the electrical power actually delivered to the datacenter.
At optimum, at most one of $E^{\text{online}}_{\text{unmet}}[k]$ and $E_\text{waste}[k]$ is nonzero.
$E^{\text{batch}}_{\text{budget}}[k]$ is the batch energy the outer planner allocates to step~$k$.
The inner scheduler spends this budget by dispatching jobs within that step.

\vspace{1.5mm}

\noindent\textbf{Putting It All Together.} Briefly, at each inner step, \sol{}'s physics update advances each module's iodine-xenon state, computes headroom, and produces the operating envelope. These per-module bounds feed into \sol{}'s outer optimizer, which re-solves every $N_\text{plan}$ buckets with fresh bounds, queue data, and grid signals.
\sol{}'s inner scheduler first serves nondeferrable demand using SMR output and grid power for any shortfall.

\makeatletter
\let\savedsection\section
\patchcmd{\section}{\@plus -2\p@ \@minus -.2\p@}{}{}{}
\makeatother
\section{Evaluation Methodology}
\let\section\savedsection
\label{sec:evaluation_methodology}

\vspace{1.5mm}
\noindent \textit{Workload and Evaluation setup.}
\sol{} is evaluated using a diverse set of benchmarks encompassing both batch and latency-sensitive online workloads.
Batch workloads span compute-bound (NPB~\cite{bailey2010parallel}, LULESH~\cite{LULESH2:changes}) and memory-bound (graph benchmarks from SeBS~\cite{copik2021sebs}) applications.

Latency-sensitive services use four TailBench~\cite{tailbench} benchmarks that operate under strict SLAs.
Job arrivals follow the Alibaba cluster trace (v2018~\cite{alibaba_clusterdata_2018}), which records co-located online and batch workloads over eight days.
The evaluation assumes a datacenter cluster of 30,000 Xeon Platinum 8175M servers (TDP of 240\,W per socket, 480\,W total CPU power).
All runtime and energy measurements were conducted on AWS m5.metal instance (96 vCPUs, 384 GB RAM), with energy measured via Likwid~\cite{psti} using RAPL~\cite{khan2018rapl} counters.
The cluster size is fixed across all policies.
The workload still reaches full machine subscription during peaks, and machine capacity is enforced through batch queueing when machine slots or power are unavailable.
Job scheduling decisions are made at one-minute granularity.
Latency-sensitive online services are served immediately, and batch jobs may be queued up to 12 hours before a deadline miss is recorded (Section~\ref{sec:scheduler_design}).

\vspace{1.5mm}
\noindent \textit{SMR plant configuration.}
At this cluster scale, analysis of the Alibaba trace reveals a mean power demand of 7\,MW and a peak offered demand of 16.1\,MW.
We therefore size the SMR plant as six modules of 1.7\,MW each, for a total SMR plant capacity of 10.2\,MW to cover the 90th percentile of demand under normal operation. During a scheduled refueling outage, the remaining five modules provide 8.5\,MW, which covers mean demand (7\,MW).
The grid interconnection, capped at one module equivalent (1.7\,MW) is a regular part of the energy mix that the scheduler actively manages.
Grid prices are from California ISO (CAISO) day-ahead wholesale prices for southern California, and grid water intensity is computed from CAISO five-minute generation mix~\cite{CAISO_OASIS_Prices,CAISO_TodaysOutlook_Supply}.
The nuclear physics parameters follow~\cite{choudhury2025physicsinformedunitcommitmentframework} and~\cite{baker2024nuscale_vera}.
Each module operates on a 22-month fuel cycle with staggered refueling (Section~\ref{sec:motivation}) and a representative 5-day outage window.

\vspace{1.5mm}
\noindent \textit{Figures of merit.}
We report batch deadline miss rate, batch average wait time, online SLA violation rate, water consumption, and grid electricity cost. Water consumption captures both the SMR's direct cooling water and the indirect water embedded in grid electricity, computed from the grid generation mix~\cite{Macknick2011_NREL_Water}.
For comparison, we evaluate three baseline configurations in which all available modules operate at full power and any shortfall is purchased from the grid, reflecting the conventional practice of operating nuclear plants as baseload sources~\cite{lokhov2021loadfollowing}. Baseline-1 is sized to mean demand (7.0\,MW), Baseline-2 to P90 demand, i.e., the 90th-percentile demand level (10.2\,MW), and Baseline-3 to peak demand (16.1\,MW).

\section{Results and Analysis}
\label{sec:results}

\begin{figure*}[t]
\centering
\includegraphics[scale=0.21]{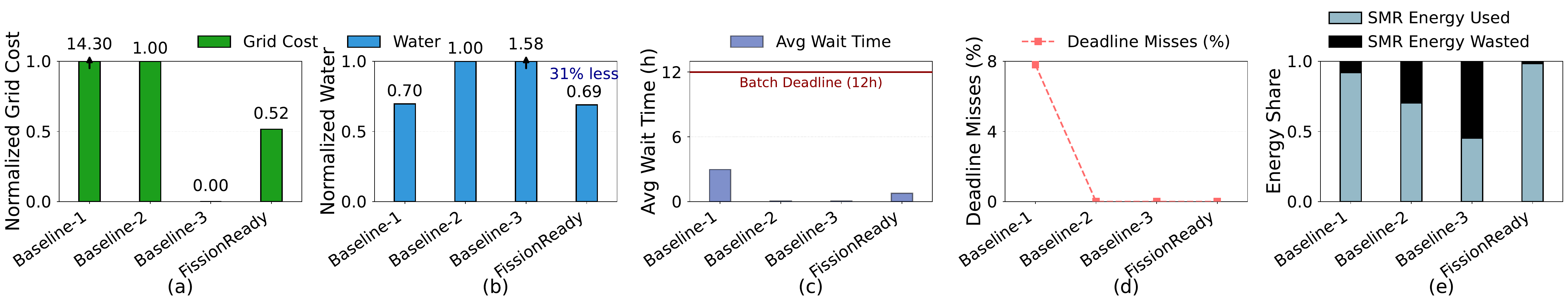}
\caption{\sol{}'s effectiveness: \sol{} reduces water by 31\%, roughly halves grid cost, preserves low batch wait with zero deadline misses, and reduces SMR energy waste compared to same-sized baseline without load following (Baseline-2).
Baseline-1 is sized to mean demand (7.0\,MW), Baseline-2 to P90 demand (10.2\,MW), and Baseline-3 to peak demand (16.1\,MW). Panels (a) and (b) are normalized to Baseline-2.
}
\label{fig:flagship2}
\end{figure*}

We present and analyze results focused on three major themes: \sol{}'s effectiveness, reasons for effectiveness, and \sol{}'s robustness to workload forecast error and reactor model mismatch .

\vspace{1.5mm}
\noindent\textbf{\sol{}'s effectiveness.} Figure~\ref{fig:flagship2} compares \sol{} against the three baseline configurations described in Section~\ref{sec:evaluation_methodology}.
Unlike various baselines, \sol{} safely load-follows datacenter demand
with zero involuntary reactor shutdowns and therefore zero deadtime and a negligible online SLA violation rate of 0.03\%.
\sol{} achieves this by reasoning about each module's fuel age, xenon concentration, and reactivity headroom to exploit the fleet's full maneuvering flexibility without pushing any module beyond its physics limits.
In comparison, Baseline-1 (7.0\,MW, mean-demand sizing) is undersized and incurs batch deadline misses (256k jobs, 7.8\%).
Baseline-3 (16.1\,MW, peak-demand sizing) reduces grid dependence but wastes thermal output that still consumes cooling water and fuel.
Baseline-2 (10.2\,MW, P90 sizing) avoids deadline misses but cannot modulate output to match demand, so excess generation is wasted whenever datacenter load falls below plant output.

At the same 10.2\,MW sizing as Baseline-2, \sol{} safely ramps individual modules to match fleet output more closely to datacenter demand.
Figure~\ref{fig:flagship2}(a), (b), and (e) show the resulting gains.
Relative to Baseline-2, \sol{} cuts grid electricity cost by roughly half, reduces water consumption by 31\%, and reduces SMR energy waste from 29.8\% to 1.7\%.
Figure~\ref{fig:flagship2}(c) and (d) show that these savings preserve workload quality of service.
\sol{} has zero batch deadline misses, an average batch wait of 0.78\,h, and its P99 batch wait is 2.32\,h, well below the 12\,h deadline.
We next examine how these scheduling decisions unfold over time and why per-module state tracking is the key enabler.

\begin{figure*}[t]
\centering
\includegraphics[width=0.56\textwidth]{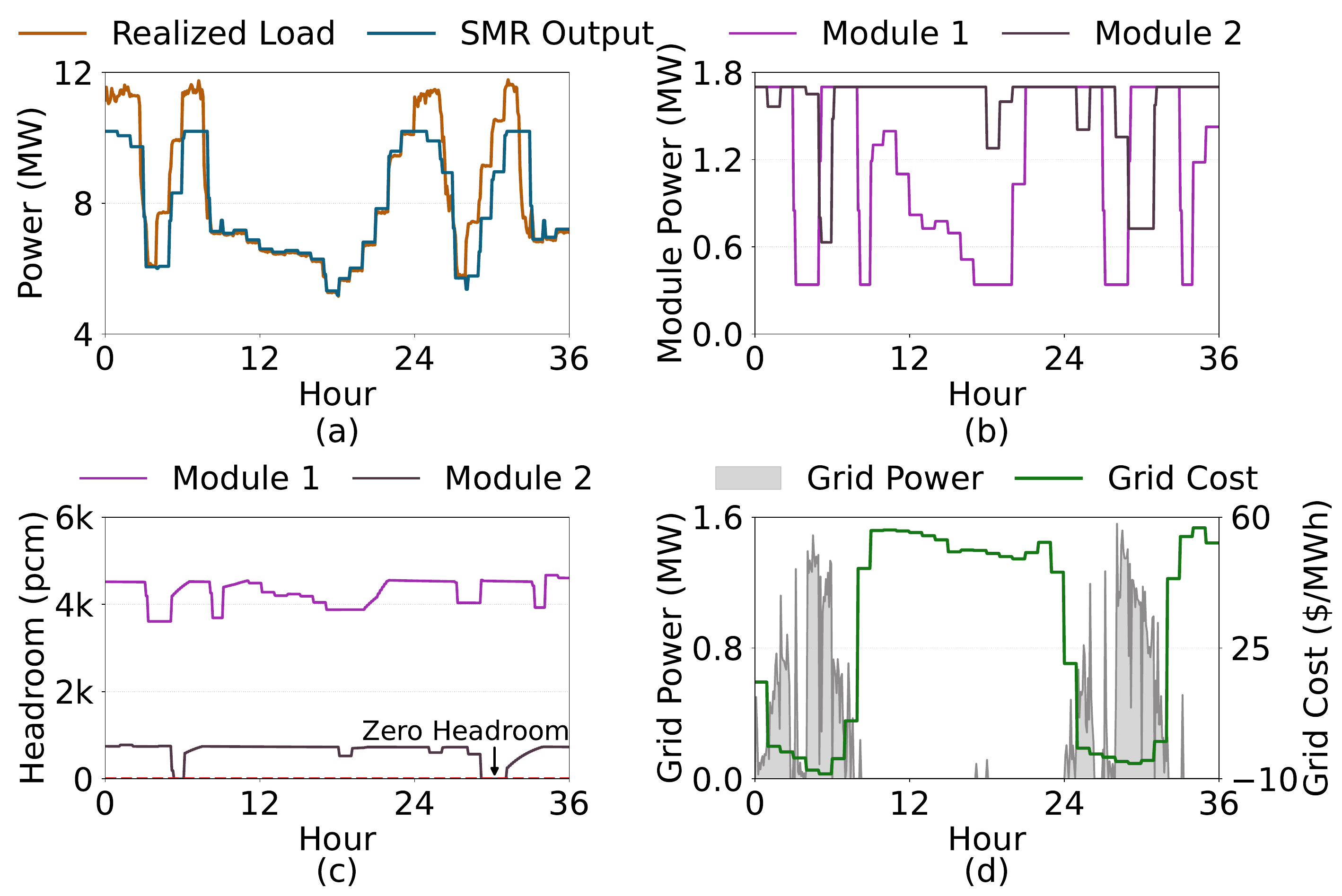}
\caption{\sol{} in action over a representative 36-hour slice from the 10.2\,MW run. \sol{} shapes the datacenter's realized load in (a), with SMR output closely tracking it, reallocates module ramps based on computed per-module headroom in (b) and (c) (module~1 is early in its fuel cycle and module~2 is near the end of cycle), and opportunistically purchases grid power during favorable cost hours in (d).
}
\label{fig:xenoflex_in_action}
\end{figure*}

\vspace{1.5mm}
\noindent\textbf{\sol{} in Action.} Figure~\ref{fig:xenoflex_in_action} shows a representative 36-hour slice from the 10.2\,MW configuration. \sol{} shapes the datacenter's realized load and keeps SMR output closely aligned with it over time. Two modules are highlighted to illustrate why per-module state tracking matters. Module~1 is early in its fuel cycle with substantial reactivity headroom, while Module~2 is near end of cycle with thinner margin.

Figures~\ref{fig:xenoflex_in_action}(b) and~(c) show that around hour 5 of the plotted window, Module~2 is ramped down from about 1.65\,MW to 0.80\,MW and then to 0.63\,MW. Over the same interval, its headroom collapses from about 750\,pcm to near zero, showing how a recent ramp-down can rapidly consume the remaining turndown margin through the delayed xenon transient.
\sol{} anticipates this evolving margin when the ramp is issued.
Once Module~2's margin is exhausted, \sol{} stops pushing it lower and shifts subsequent flexibility to the less constrained module.
Around hours 8.0 to 8.2, for example, Module~1 is reduced from 0.85\,MW to 0.34\,MW while Module~2 remains at 1.7\,MW, reflecting the much larger headroom available on Module~1.

Figure~\ref{fig:xenoflex_in_action}(d) shows that \sol{} purchases grid power during favorable cost and water hours.
During this window, purchased grid power averages $-\$1.5$/MWh, compared to $\$28.5$/MWh over the full window, and it comes from lower-water grid hours on average (346 versus 475~L/MWh).
Through these coordinated ramping, load-shaping, and grid-purchase decisions, \sol{} translates its scheduling choices into system-level savings.
A natural question is whether per-module physics tracking is necessary for these gains, or whether one fixed module power range for all modules could suffice.

\vspace{1.5mm}
\noindent\textbf{Why Per-Module State Tracking Matters.} To answer this question, we compare against Fixed Load Following (F-LF). It uses the same optimization framework as \sol{}, but replaces \sol{}'s time-varying per-module state tracking with a uniform load-following range that applies to all modules.
Without tracking per-module xenon and iodine concentrations, a module can be commanded to ramp beyond what its current state can safely sustain.
When xenon reactivity exceeds the compensation ceiling, the module can no longer maintain criticality and shuts down involuntarily.
The module is then stranded offline until the xenon transient decays on its own timescale, which can take 20 hours or more near end of cycle.
Unplanned shutdowns are among the serious operational events at a nuclear plant, carrying both safety and economic consequences.

\begin{table*}[t]
 \centering
 \caption{\sol{}'s load following outperforms both a Fixed Load Following baseline and the day-ahead DCSMR baseline at the same 10.2\,MW sizing.}
 \label{tab:flf_comparison}
 \begingroup
 \setlength{\tabcolsep}{5pt}
 \scalebox{0.95}{
 \begin{tabular}{lccccccc}
 \toprule
 & \makecell[c]{\textbf{}\\\textbf{Shutdowns}} & \makecell[c]{\textbf{Deadtime}\\\textbf{(h)}} & \makecell[c]{\textbf{Average Wait}\\\textbf{Time (h)}} & \makecell[c]{\textbf{SLA}\\\textbf{Violations (\%)}}
& \makecell[c]{\textbf{Deadline}\\\textbf{Misses (\%)}} & \makecell[c]{\textbf{Grid}\\\textbf{Cost}} & \makecell[c]{\textbf{}\\\textbf{Water}} \\
 \midrule
 Fixed Load Following & 11 & 212 & 0.26 & 0.16 & 0.0 & 30.5 & 0.91 \\
 DCSMR & 6 & 127 & 7.86 & 2.24 & 0.06 & 9.4 & 0.96 \\
 \sol{} & 0 & 0 & 0.78 & 0.03 & 0.0 & 1.0 & 1.0 \\
 \bottomrule
 \end{tabular}
 }
 \endgroup
\end{table*}

Table~\ref{tab:flf_comparison} summarizes the comparison. F-LF causes 11 involuntary shutdowns with 212 hours of cumulative deadtime, multiple days of lost module capacity. In contrast, \sol{} achieves zero shutdowns and zero deadtime because its formulation identifies which modules can safely absorb a ramp and by how much before committing to it.
The cascading effects of these shutdowns are visible across every metric: F-LF's grid costs are 30$\times$ higher because it purchases grid power reactively during unplanned outages at unfavorable prices, and it incurs nearly 5$\times$ more SLA violations because unexpected module outages leave the fleet unable to meet online demand.

\vspace{1.5mm}

We also adapt the aggregate load-following SMR scheduler (DCSMR) of Yang et al.~\cite{yang2025dcsmr} to the same workload, queueing, grid signals, and objective setting as \sol{}.
DCSMR treats the SMR fleet as a single aggregate source and plans its output over a single day-ahead horizon, enforcing aggregate ramp-rate and stable-period constraints after ramp-down.

F-LF and DCSMR achieve lower absolute water consumption (9.3 and 9.8 versus 10.2 million liters), but that reduction is an artifact of lost output.

\noindent\textit{Scheduling overhead.} \sol{}'s optimizer incurs only 35\,ms overhead on average for a 6-module fleet with a 48-hour horizon -- less than 0.006\% of each 10-minute scheduling interval (overhead dependent on number of modules, instead of cluster size).

\vspace{1.5mm}
\noindent\textbf{\sol{}'s Robustness and Sensitivity.}

\begin{table}[t]
\centering
\caption{Robustness under reactor-model misalignment and forecast noise.
(a) Involuntary shutdowns across xenon destruction perturbations and safety reserves.
The benign case removes xenon faster than predicted, while the remaining perturbations remove it more slowly.
(b) Sensitivity to workload forecast noise. Water savings in both panels are relative to SMR without load following (Baseline-2), the fixed-output SMR fleet sized to P90 demand (10.2\,MW).}
\label{tab:robustness_sweep_tables}
{
\begin{minipage}{\linewidth}
\centering
\textbf{(a) Reactor-Model Misalignment}\\[2mm]
\setlength{\tabcolsep}{4.0pt}
\resizebox{0.8\linewidth}{!}{
\begin{tabular}{lccccc}
\toprule
\multirow{2}{*}{\makecell[l]{Reactor-model\\misalignment}} & \multicolumn{5}{c}{$ \text{safety reserve} (\sigma_\text{margin})$ (pcm)} \\
\cmidrule(lr){2-6}
 & 0 & 150 & 300 & 450 & 600 \\
\midrule
Benign $+5\%$ & 0 & 0 & 0 & 0 & 0 \\
Default & 0 & 0 & 0 & 0 & 0\\
$5\%$ & 34 & 0 & 0 & 0 & 0 \\
$10\%$ & 45 & 1 & 0 & 0 & 0 \\
$15\%$ & 36 & 24 & 0 & 0 & 0 \\
$20\%$ & 42 & 26 & 3 & 0 & 0 \\
$25\%$ & 50 & 43 & 28 & 3 & 0 \\
Water Saved (\%) & 30.3 & 29.8 & 28.2 & 27.0 & 25.5\\
\bottomrule
\end{tabular}
}
\end{minipage}\par\medskip
\begin{minipage}{\linewidth}
\centering
\textbf{(b) Forecast Quality}\\[2mm]
\setlength{\tabcolsep}{4.0pt}
\resizebox{0.8\linewidth}{!}{
\begin{tabular}{lccccc}
\toprule
Forecast & \makecell[c]{Water\\Saved (\%)} & \makecell[c]{Wait\\(h)} & \makecell[c]{Deadline\\Miss (\%)} & \makecell[c]{SLA\\(\%)} & Shutdowns \\
\midrule
Default & 31.0 & 0.78 & 0.0 & 0.03 & 0 \\
Noisy 10\% & 30.6 & 0.81 & 0.0 & 0.06 & 0 \\
Noisy 25\% & 29.6 & 0.96 & 0.0 & 0.34 & 0 \\
Noisy 50\% & 28.8 & 1.31 & 0.0 & 1.74 & 0 \\
\bottomrule
\end{tabular}
}
\end{minipage}
}

\end{table}

\vspace{1.5mm}
\noindent\textit{Robustness to misalignment between the operating reactor and underlying reactor model.} Recall that \sol{} has two components: a nonlinear physics component with a xenon-iodine reactor model (the reactor model), and a scheduling optimization component, as described in Section~\ref{sec:optimization_approach}. The scheduler component interacts with the reactor model through per-module power bounds produced by solving the xenon-iodine dynamics.

The reactor models are calibrated periodically and always have sufficient safety margins. Between calibration periods, the operating reactor may develop a small transient misalignment from the reactor model. Therefore, we evaluate whether this transient mismatch affects the scheduler’s effectiveness (involuntary shutdowns and water savings).

To evaluate this, we artificially perturb the xenon destruction dynamics in the operating reactor so that it drifts away from the underlying reactor model -- essentially emulating as if the operating reactor and reactor model were not aligned. This artificial perturbation removes xenon more slowly after a power reduction, so a ramp can produce a higher xenon peak and consume more headroom than expected by the scheduler component. To be even more pessimistic, we perform this sensitivity on an aged fleet, where reactivity headroom is the smallest. Table~\ref{tab:robustness_sweep_tables}(a) shows the resulting involuntary shutdowns for a range of perturbations and safety reserves (since increasing safety reserve reduces involuntary shutdowns).

Overall, we highlight major takeaways from our results (Table~\ref{tab:robustness_sweep_tables}(a)). First, even the transient misalignment between the operating reactor and reactor model does not lead to involuntary shutdowns until a significant 15\% perturbation to the xenon-destruction dynamics. Only beyond 15\% misalignment, we observe involuntary shutdowns. Second, even when the misalignment is more than 15\%, fortunately, one can limit or eliminate the involuntary shutdowns by increasing the safety reserve (beyond the default value of 300\,pcm).

\noindent\textit{Robustness to workload forecast error.}
\sol{} uses a simple workload forecaster to estimate future nondeferrable demand and batch arrivals.

 We evaluate robustness to forecast error by applying 10\%, 25\%, and 50\% multiplicative noise to both forecasts before each planning solve.

At dispatch time, each module remains constrained by its current physics envelope, independently of the workload forecast.

The inner dispatcher observes the actual queue at each step, and the outer planner updates the remaining plan at the next replan.
Table~\ref{tab:robustness_sweep_tables}(b) shows that 10\% forecast noise changes little.
Even at 25\% noise, average batch wait rises from 0.78\,h to 0.96\,h and the online SLA violation rate rises from 0.03\% to 0.34\%, while deadline misses and involuntary shutdowns remain zero.

\vspace{1.5mm}
\noindent\textit{Year-long operation.}
To evaluate whether the monthly results persist through reactor state evolution, refueling, workload variation, and seasonal grid conditions, we evaluate \sol{} and Baseline-2 continuously for one year.
The SMR plant configuration remains unchanged, and we use a full year of CAISO price and generation mix data.
Compared with Baseline-2, \sol{} reduces annual water use from 211 million liters to 139 million liters, a 34.1 percent reduction, and cuts SMR energy waste from 31.3 to 1.6 percent.
Average batch wait is 0.80\,h, the batch deadline miss rate is zero, and the online SLA violation rate is 0.07 percent.
The fleet absorbs three staggered refueling outages without a deadline miss.
The fleet ends the year with zero involuntary shutdowns and zero deadtime.

\section{Related Work}
\label{sec:related_work}

\noindent\textbf{Nuclear-powered datacenters and SMR flexibility.} At the power-system level, flexible nuclear operation can lower system costs and reduce renewable curtailment~\cite{jenkins2018nuclear}.
Some early works have explored potential from the standpoint of technical feasibility, economics, and plant operation~\cite{lokhov2021loadfollowing,zhang2024load}.
Prior work has primarily investigated plant dynamics and component wear under flexible operation~\cite{baker2024nuscale_vera,holos2022load}, reactor-level load-follow control and flexible-operation economics~\cite{franceschini2008mshim,ponciroli2017profitability}, ramp-rate and stable-period constraints~\cite{yang2025dcsmr}, multi-unit SMR dispatch at the power-system level~\cite{alhadhrami2023dispatch}, and fuel-cycle-awareness in unit-commitment formulations~\cite{choudhury2025physicsinformedunitcommitmentframework}.

Recent interest in nuclear-powered datacenters has focused on siting, sizing, and coupling scenarios for co-located nuclear and digital infrastructure~\cite{stauff2025nuclearpowered}, but fine-grained module-aware schedulers for SMR-powered datacenter operations remain largely unexplored.
Therefore, \sol{} is naturally timely, critical, and unique in multiple ways.
First, this is a novel job scheduler for emerging SMR-powered data centers that have multi-module SMR plants with staggered refueling.
This naturally captures temporary module unavailability during refueling outages.
Second, it tracks per-module fuel age together with iodine-xenon state to compute module-specific power ranges and headroom before each replanning step.
Third, \sol{} is the first solution to couple physics abstraction to a two-timescale scheduler that plans fleet output and grid use over a horizon while making finer-grained online dispatch decisions, jointly accounting for batch deadlines, online-service quality of service, grid price, and water consumption.
To the best of our knowledge, \sol{} is the first datacenter scheduler to integrate per-module xenon-aware SMR load following with water-aware workload shaping.

\vspace{1mm}
\noindent\textbf{Energy-, carbon-, and water-aware datacenter scheduling.}
Datacenter systems work has long exploited workload slack to follow external energy signals.
Temporal shifting approaches defer batch jobs to periods of low carbon intensity~\cite{radovanovic2021carbonawarecomputingdatacenters,wiesner2021waitawhile,hanafy2023carbonscaler}, while spatial approaches migrate workloads across geo-distributed data centers~\cite{gsteiger2024caribou}.
For example, Ecovisor virtualizes energy to provide per-application carbon budgets~\cite{souza2023ecovisor}, and Carbon Explorer offers a holistic framework combining renewables, batteries, and scheduling~\cite{acun2023carbonexplorer}.
Water as a resource is beginning to attract more attention; recent works co-optimize water consumption, carbon emissions, and electricity cost through geographic load balancing across distributed sites~\cite{islam2018wace,yankai2024waterwise,moore2025slit,islam2015watch}.
These works treat the energy supply as externally given and do not address the water-consumption burden of SMRs.
\sol{} instead schedules against a controllable on-site SMR whose feasible flexibility depends on internal reactor state, and it jointly optimizes workload quality of service, grid cost, and water consumption.

\section{Conclusion}
\label{sec:conclusion}
\sol{} is a scheduler for SMR-powered datacenters that couples per-module reactor physics with datacenter workload management.
\sol{} tracks fuel age and iodine-xenon dynamics at the module level, exposes the resulting maneuvering flexibility through reactivity headroom, and combines this physics abstraction with two-timescale control of fleet output and workload dispatch.
In the main and year-long evaluations, \sol{} achieves substantial water and grid-cost savings while preserving workload quality of service and completely avoiding involuntary shutdowns.
More broadly, \sol{} is a case study in the unique challenges and opportunities of scheduling for SMR-powered datacenters, and it highlights the need for future work in this space to jointly consider physics, workload, and grid dynamics.

\bibliographystyle{ACM-Reference-Format}
\bibliography{refs}

\end{document}